# DC Microgrid based on Battery, Photovoltaic, and fuel Cells; Design and Control

Akram Muntaser[1], Abdurazag Saide[1], Hussin Ragb[2], and Ibrahim Elwarfalli[3]
[1]University of Dayton, emails: muntasera1@udayton.edu, saidea1@udayton.edu
[2]Christian Brothers University, email: hragb@cbu.edu
[3]West Virginia University, email: ieelwarfalli@mix.wvu.edu

**Abstract:**

Microgrids offer flexibility in power generation in a way of using multiple renewable energy sources. In the past few years, microgrids become a very active research area in terms of design and control strategies. Most of the microgrids use DC/DC converters to connect renewable energy sources to the load. In this paper, the simulation model of a DC microgrid with three different energy sources (Lithium-ion battery (LIB), photovoltaic (PV) array, and fuel cell) and external variant power load is built with MATLAB/Simulink and the simulative results show that the stability of DC microgrid can be guaranteed by the proposed maximum power point controller MPPT. The three energy sources are connected to the load through DC/DC converters, one for each. This type of topology ensures protection for each energy source as well as optimum stability at the load.

## I.  INTRODUCTION

Renewable energy nowadays is 19% of the global power generation as shown in Fig.1. Recently Microgrid has been rapidly developing to reduce environmental pollution and increase the consumption of renewable energy. A microgrid is a system composed of distributed generations, energy storage systems, power electronic converters, loads, and energy management systems [1,2]. Due to the advantages of simple structure, flexible control strategies, simple energy conversion, and high efficiency [3,4]. In addition, DC Microgrids are attracting more and more attention from all over the world, most of renewable sources generate DC and require an additional conversion step before interconnection to the AC grid. A DC Microgrid has many advantages over AC Microgrid, because it needs only few power converters with higher system efficiency and easier interface of renewable energy sources to a DC System [5]

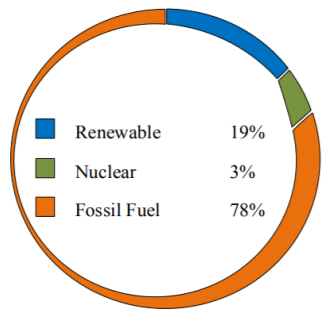

Fig.1. Global energy production

A microgrid is an active power distribution network, which has the capability of autonomous operation. The essential components of a microgrid are distributed generators (DG), energy storage elements, and controllable loads [6, 7]. The unique advantage of a microgrid is its ability to operate both in grid-connected and islanded (or autonomous) modes. Microgrids can be classified as AC microgrids and DC microgrids depending on the nature of bus voltage [8]. In an AC microgrid, the distributed generators are connected to the AC bus using power electronic converters and the alternating current (AC) loads are directly connected to the AC bus. AC microgrids are more popular compared to DC microgrids since the existing power distribution networks are predominantly AC-based. In AC microgrids, the power produced by renewable energy sources is fed to the grid after synchronizing the voltage with the grid voltage to operate in grid-connected mode. Also in islanded mode, the control of the DG's, loads, and energy storage equipment to maintain a stable voltage is very complicated.

Recently too much research has focused on DC microgrids since DC microgrids have several advantages over AC microgrids. Some of the renewable energy sources such as solar and fuel cells produce DC power which is suitable for most of the existing equipment and devices such as computers, phones, LED lamps, and even electric vehicles work on DC power, DC microgrid presents itself as a more feasible alternative over AC microgrid. The potential merits of DC microgrids over AC microgrids are given as: I- the overall efficiency is improved as the unnecessary AC/DC power conversions are reduced. II- Simple and cost-effective power electronics interfaces to connect the sources and loads to the microgrid bus. III- No issues of reactive power flow and easier integration of energy storage devices [9]. As mentioned, most of microgrids use DC/DC converters to connect the renewable energy sources to the load. DC/DC converters have been widely used in distributed power generation systems [10,11], electric vehicles [12,13] and uninterruptible power supply systems, and other emerging energy conversion systems. With the increasing use of DC micro-power and DC load, DC microgrids with energy storage systems have broad development prospects [14].

In this paper, the methodology of the system including the basic concepts of the DC microgrid architecture and system configuration is discussed in section I along with the fundamental theory of the system components (renewable energy sources and D/DC converters). Section II discusses the control design and implementation. Section III includes the simulated results obtained with the Maximum Power Point Tracking controller (MPPT) strategy. Section IV concludes the work in this paper with results discussion and future proposed work.

## I. SYSTEM DESIGN AND COMPONENTS

In this paper, we introduce a proposed microgrid system with three different energy sources LIB, PV array, and fuel cells, and controlled using a MPPT controller. The three different energy sources are connected to DC/DC converters which are connected to the load. This type of topology ensures that each source has its own protector converter which fits its different characteristics. Fig.2 shows the block diagram of the system.

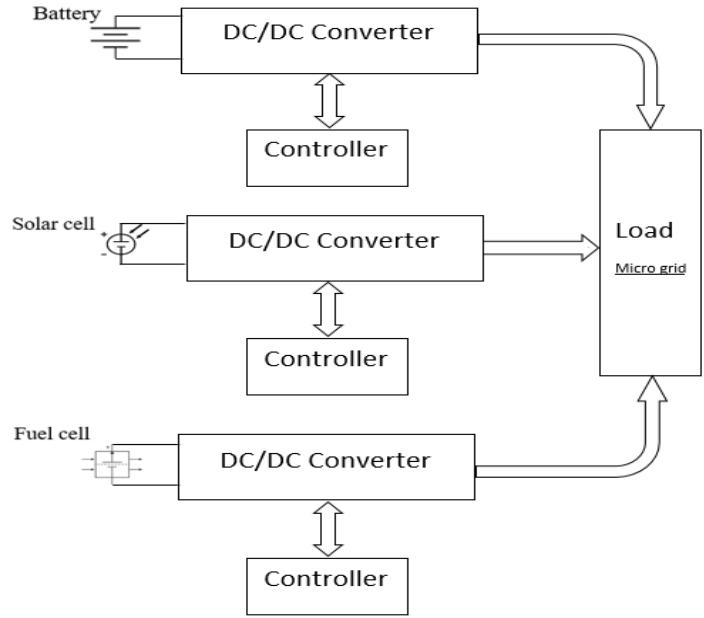

Fig.2. Block diagram of the system

*Lithium-ion battery*

Lithium-ion battery (LIB) is the most common type of batteries commercially used these days and that is due to its features such as high energy density, lack of memory effect, and high charge and discharge rate capabilities [15,16]. The equivalent circuit of the battery is shown below in Fig.3:

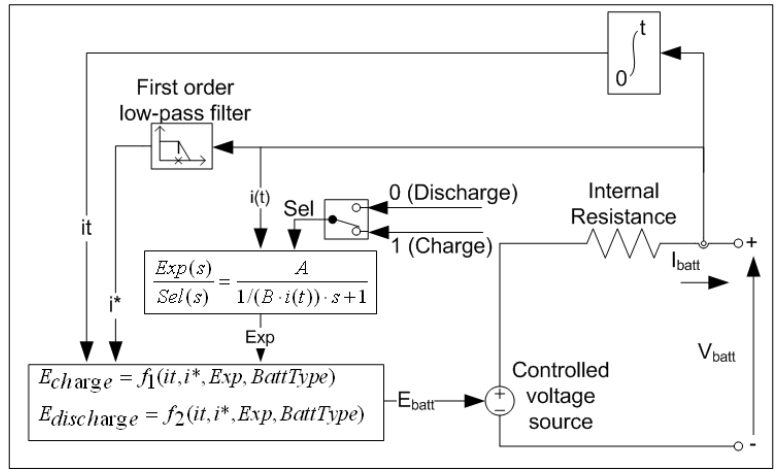

Fig.3. Battery equivalent circuit

Whereas,

$E_{Batt}$ is the nonlinear voltage (V)

$E_0$ is the constant voltage (V)

$Exp$ is the exponential zone dynamics (V)

$Sel$ represents the battery mode.

$Sel = 0$ during battery discharge, $Sel = 1$ during battery charging.

$K$ is the polarization constant (Ah$^{-1}$) or Polarization resistance (Ohms)

$i*$ is the low-frequency current dynamics (A)

$i$ is the battery current (A)

$it$ is the extracted capacity (Ah)

$Q$ is the maximum battery capacity (Ah)

$A$ is the exponential voltage (V)

$B$ is the exponential capacity (Ah), [14].

The nominal voltage (V) represents the end of the linear zone of the discharge characteristics while the rated capacity is the minimum effective capacity of the battery. The initial State-Of-Charge (SOC) of the battery is 100% indicating a fully charged battery. These parameters are used as an initial condition for the system. The internal resistance of the battery (ohms) is supposed to be constant during the charge and the discharge cycles and does not vary with the amplitude of the current.

*Solar cell (PV Array)*

A solar cell is an electronic device which directly converts sunlight into electricity. Light shining on the solar cell produces both a current and a voltage to generate electric power. This process requires firstly, a material in which the absorption of light raises an electron to a higher energy state, and secondly, the movement of this higher energy electron from the solar cell into an external circuit. The electron dissipates its energy in the external circuit and returns to the solar cell. A variety of materials and processes can potentially satisfy the requirements for photovoltaic energy conversion, but in practice, nearly all photovoltaic energy conversion uses semiconductor materials in the form of a p-n junction. Regarding the development of sustainable energy, such as solar energy, Fig.4 shows a generic solar cell.

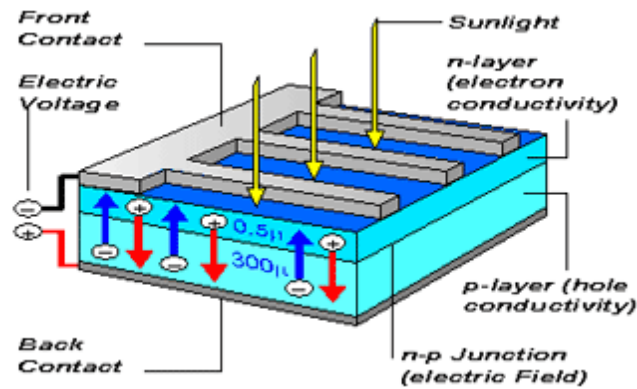

Fig.4. Solar cell

In our design, we used the PV array model, which implements an array of PV built of strings of modules connected in parallel, each string consisting of modules connected in series. The PV Array block is a five-parameter model using a current source $I_l$ (light-generated current), diode ($I_d$ and other parameters), series resistance $R_s$, and shunt resistance $R_{sh}$ to represent the irradiance- and temperature-dependent I-V characteristics of the modules. Fig.5 shows the PV Array equivalent circuit.

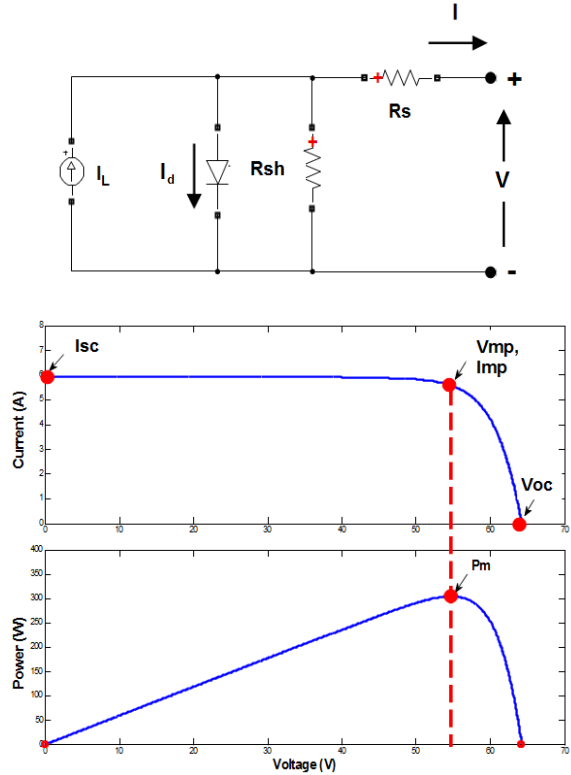

Fig.5. PV array equivalent circuit and characteristics

The diode I-V characteristics for a single module are defined by the equations

$$I_d = I_o \left[ exp\left(\frac{V_d}{V_T}\right) - 1 \right] \quad (1)$$

$$V_T = \frac{KT}{q} \times nI \times Ncell \quad (2)$$

*Whereas*:

$I_d$ is the diode current (A)

$V_d$ is the diode voltage (V)

$I_o$ is the diode saturation current (A)

$nI$ is the diode ideality factor, a number close to 1.0

$k$ is the Boltzman constant = 1.3806e-23 J.K-1

$q$ is the electron charge = 1.6022e-19 C

$T$ is the cell temperature (K)

$N_{cell}$ is the number of cells connected in series in a module

In our design, we considered a 6-kW PV array that uses 330 sun power modules. The array consists of 66 strings of 5 series-connected modules connected in parallel (10*2*305.2 W= 6.1 kW).

*Fuel cell*

A fuel cell is an electrochemical device that produces electricity without combustion by combining hydrogen and oxygen to produce water and heat. A fuel cell is a device that uses hydrogen (or hydrogen-rich fuel) and oxygen to create electricity by an electrochemical process. A single fuel cell consists of an electrolyte squeezed in between two thin electrodes (a porous anode and cathode) Hydrogen, or a hydrogen-rich fuel, is fed to the anode where a catalyst separates hydrogen's negatively charged electrons from positively charged ions (protons) At the cathode, oxygen combines with electrons and, in some cases, with species such as protons or water, resulting in water or hydroxide ions, respectively The electrons from the anode side of the cell cannot pass through the membrane to the positively charged cathode; they must travel around it via an electrical circuit to reach the other side of the cell. This movement of electrons is an electrical current. The amount of power produced by a fuel cell depends upon several factors, such as fuel cell type, cell size, the temperature at which it operates, and the pressure at which the gases are supplied to the cell. Fig.6 shows a generic fuel cell.

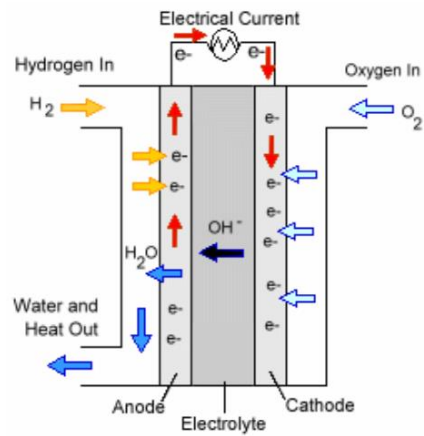

Fig.6. Fuel cell

In our design, we used the fuel cell stack model which implements a generic model parameterized to represent the most popular types of fuel cell stacks fed with hydrogen and air. This model is based on the equivalent circuit of a fuel cell stack shown in Fig.7:

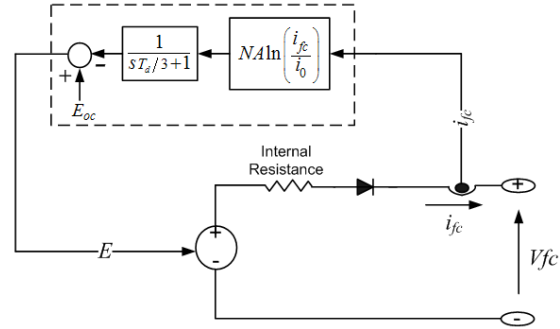

Fig.7. Fuel cell equivalent circuit

This model represents a particular fuel cell stack operating at nominal conditions of temperature and pressure. The parameters of the equivalent circuit can be modified based on the polarization curve obtained from the manufacturer datasheet. A diode is used to prevent the flow of negative current into the stack. A typical polarization curve consists of three regions:

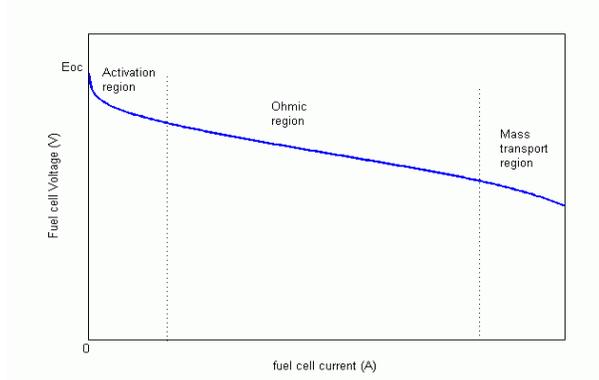

Fig.8. Fuel cell polarization curve

The first region represents the activation voltage drop due to the slowness of the chemical reactions taking place at electrode surfaces. Depending on the temperature and operating pressure, type of electrode, and catalyst used, this region is more or less wide. The second region represents the resistive losses due to the internal resistance of the fuel cell stack. The third region represents the mass transport losses resulting from the change in concentration of reactants as the fuel is used.

*DC/DC converter (buck)*

Several structures of bidirectional DC/DC converters and control strategies of energy storage systems have been studied in [17,18]. Buck DC/DC converter is used to step the power down and regulate it in a certain desired level. In this system, the buck converter has been used to regulate and control the LIB power and provide it to the load. DC/DC buck converter basically is a powerful tool that is used to step down the higher voltage to a lower voltage. Buck converter consists of LC low bass filter to regulate the source voltage and lowered down as desired. The source will be connected to switch works using a duty cycle in order to open and close in friction of second. It can be any type of switches but for more efficiency, a MOSFET switch has been used. The reverse bias diode is used for circuit protection, so the current would have a path to go through whenever the MOSFET is open. Fig.9 shows a simple circuit for the buck converter.

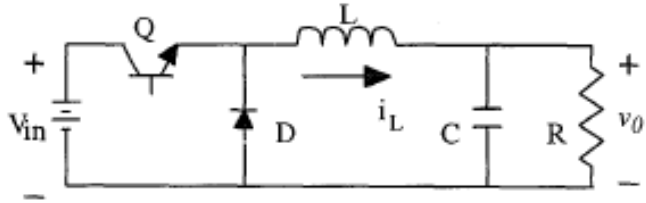

Fig.9. Buck converters

The buck converter with ideal switching devices will be considered here which is operating with the switching period of T and duty cycle D. The state equations corresponding to the converter in continuous conduction mode (CCM) can be easily understood by applying Kirchhoff's voltage law on the loop containing the inductor and Kirchhoff's current law on the node with the capacitor branch connected to it. When the ideal switch is ON, the dynamics of the inductor current $i_L(t)$ and the capacitor voltage $V_c(t)$ are given by,

$$\begin{cases} \dfrac{di_L}{dt} = \dfrac{1}{L}(V_{in} - v_o) \\ \dfrac{dv_o}{dt} = \dfrac{1}{C}(i_L - \dfrac{v_o}{R}) \end{cases}, \quad 0 < t < dT, \quad Q:ON$$

When the switch is OFF are presented by,

$$\begin{cases} \dfrac{di_L}{dt} = \dfrac{1}{L}(-v_o) \\ \dfrac{dv_o}{dt} = \dfrac{1}{C}(i_L - \dfrac{v_o}{R}) \end{cases}, \quad dT < t < T, \quad Q:OFF$$

- **DC/DC converter (boost)**

DC/DC Boost converter simply consists of inductor, capacitor, diode, and switch. Circuit is connected as showing in Fig.10. The IGBT switch works using high frequency duty cycle in order to open and close in friction of second. The switch can be any kind of switches, but IGBT has been used in order to increase the efficiency. Boost converter works as following: when the switch is open, current flow through the circuit and charge the output capacitor using input voltage. When the switch closed, current would flow through the small, short circuit, and inductor will store the energy in form of magnetic field, then when the switch opened again, this energy will add up with source energy to charge the output using higher current. Since the MOSFET switch open and close in friction of second, that would not affect the coil.

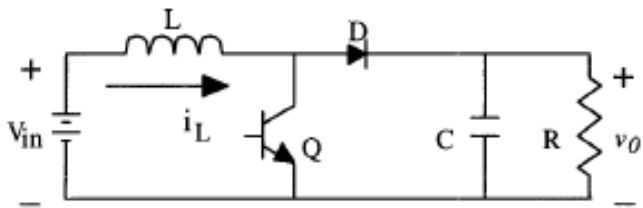

Fig.10. boost converters

The boost converter with a switching period of T and a duty cycle of D is given. Again, assuming continuous conduction mode of operation, the state space equations when the main switch is ON are shown by,

$$\begin{cases} \dfrac{di_L}{dt} = \dfrac{1}{L}(V_{in}) \\ \dfrac{dv_o}{dt} = \dfrac{1}{C}\left(-\dfrac{v_o}{R}\right) \end{cases}, \quad 0 < t < dT, \quad Q:ON$$

When the switch is OFF

$$\begin{cases} \dfrac{di_L}{dt} = \dfrac{1}{L}(V_{in} - v_o) \\ \dfrac{dv_o}{dt} = \dfrac{1}{C}\left(i_L - \dfrac{v_o}{R}\right) \end{cases}, \quad dT < t < T, \quad Q:OFF$$

## II. CONTROLLER DESIGN

Many conventional [19] and robust controllers [20,21] has been used to control the stability and robustness of microgrid. PID and Fuzzy logic are popular used controllers for microgrid and other applications [22]. In this paper, Maximum Power Point Tracking controller (MPPT) which is an algorithm that included a controllers used for extracting maximum available power from the microgrid module under certain conditions. The flowchart of this algorithm with PV array source is presented in Fig.11.

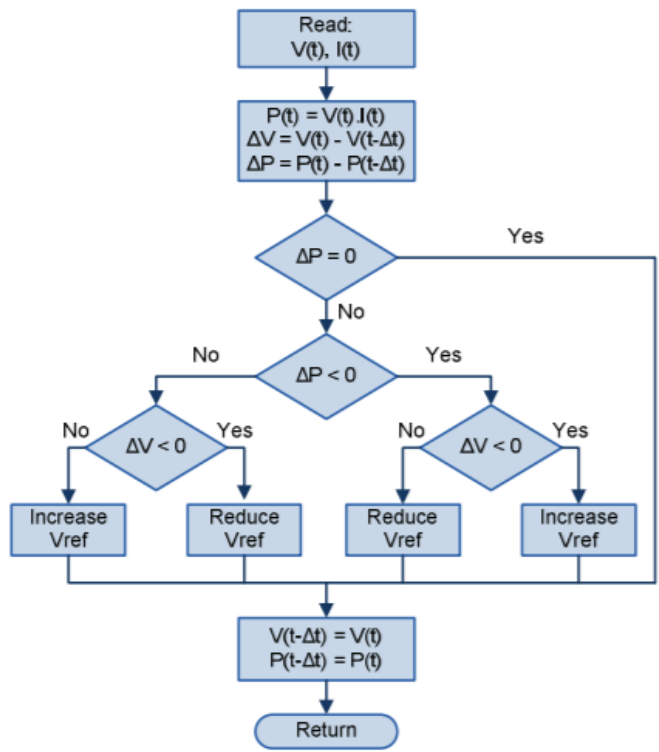

Fig.11. Flowchart of the P&O Algorithm.

The voltage which microgrid module can produce maximum power is called 'maximum power point'. Maximum power also varies with solar radiation and solar cell temperature. The work principle of MPPT is to extract the maximum available power from microgrid module by making them operate at the most efficient voltage (maximum power point). By other words, the controller checks the output of the energy source, compares it to battery and/or solar cell voltage then fixes what is the best power that microgrid module can produce to charge the energy source and converts it to the best voltage to get maximum current into load.

In this design, we applied the Perturb and Observe (P&O) MPPT controller, The P&O method operates periodically incrementing or decrementing the output terminal voltage of the PV and comparing the power obtained in the current cycle with the power of the previous cycle. If the voltage varies and the power increases, the control system changes the operating point in that direction, otherwise change the operating point in the opposite direction. Once the direction for the change of current is known, the current is varied at a constant rate. This rate is a parameter that should be adjusted to allow the balance between faster responses with less fluctuation in steady state [23].

### III. EXPERIMENTAL RESULTS

The main goal of our DC Microgrid design is to produce the required power using three different energy sources. As mentioned in introduction, we have three power sources (LIB, Fuel cell, and Solar cell), which they provide a certain power amount controlled and regulated using the DC/DC converters. The Microgrid have been tested using couple different reference signals. Fig.12 shows the first response which is a constant power demand with no changes.

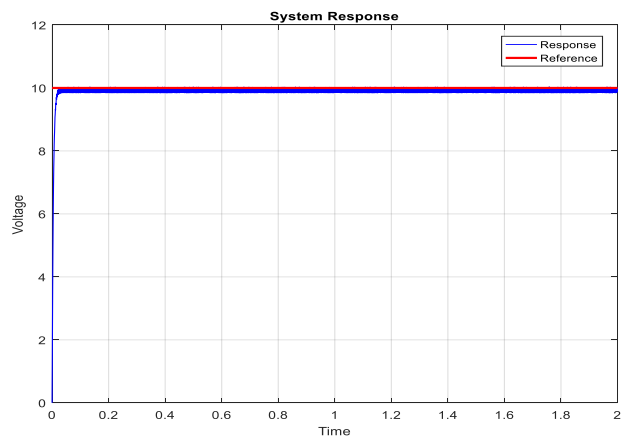

Fig.12. Microgrid response with controller

Second signal will be a power demand with a different level, which require the system to provide a high power, then a less power levels, then vice versa, to see how the MPPT controller deal with that. The system responded as expected with a very small drop off occurred to the signal when it changes from level to another, as shown in Fig.13.

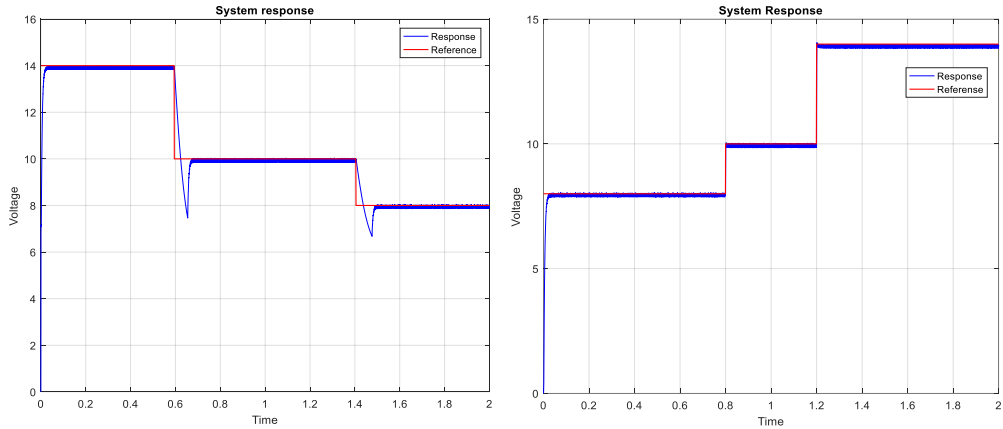

Fig.13 Microgrid response for a different power demand level

Next, we test the microgrid system with a continuously changing power demand, increasing and decreasing output. Fig.14 shows that the system responded excellent with these output demands. These power demands simulate the hybrid electrical vehicles power demand at process of speeding up and slowing down.

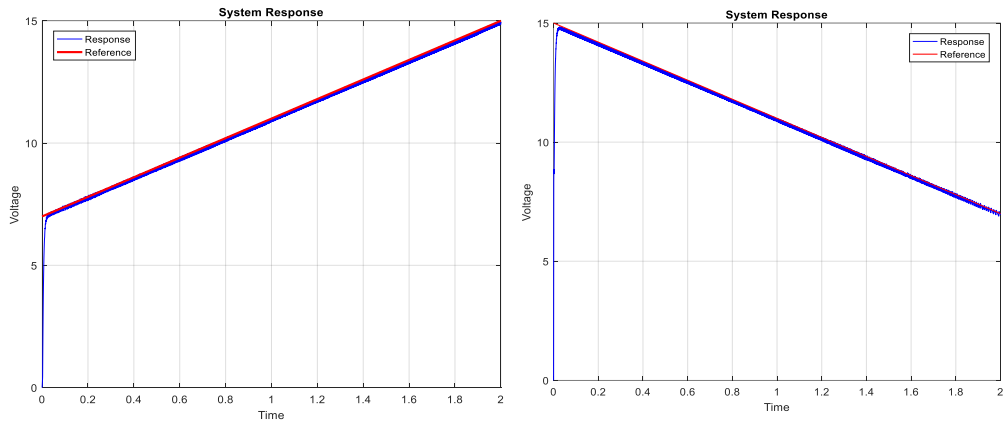

Fig.14 Microgrid response for a continuously changing power demand

The topology of connecting DC/DC converters with energy sources in parallel, allow them to share the load in proportion to their power ratings. MPPT controller method is a very effective method to ensure the optimum load sharing between parallel DC/DC converters in a DC microgrid. Since the output voltage of each DC/DC converter is separated, the terminal voltage itself provides information regarding the power delivered. Consequently, the benefit of this method is that there is no requirement of a separate communication link between the DC/DC converters. Since this method being a decentralized method without any communication link, the reliability is higher compared to centralized control methods.

## IV. Conclusion

Unpredictable variation in renewable microgrid energy sources and load demand is the biggest challenge in microgrid system design. As of the non-linear responses of the DC microgrid and considering the uncertainties and disturbances of system parameters, the MPPT controller is adopted to control the DC bus voltage and track the demanded output voltage quickly to ensure the robustness and stability of the system. The validity and suitability of the proposed control technique with DC/DC conversion of a bidirectional DC microgrid system are validated by simulation. The control strategy is simple in algorithm and has good reference value for large-scale stability control of power converter operating point.